%% file: Manuscript.tex
\documentclass[conference]{IEEEtran}
\IEEEoverridecommandlockouts
\usepackage{cite}
\usepackage{amsmath,amssymb,amsfonts}
\usepackage{graphicx}
\usepackage{textcomp}
\usepackage[ruled]{algorithm2e}
\usepackage{algpseudocode}
\usepackage{xcolor}
\usepackage{braket}
\usepackage{multirow}
\usepackage{subfigure}
\usepackage{blkarray}
\usepackage{bm}

\usepackage[switch,mathlines,displaymath]{lineno}

\usepackage{etoolbox}
\makeatletter
\patchcmd{\@makecaption}
  {\scshape}
  {}
  {}
  {}
\makeatletter
\patchcmd{\@makecaption}
  {\\}
  {.\ }
  {}
  {}
\makeatother

\usepackage{caption}

\usepackage[normalem]{ulem}

\newcommand{\SVD}{{\rm SVD}}

\newcommand{\shang}[1]{{{#1}}}  
\newcommand{\fanyi}[1]{{{#1}}}
\newcommand{\liu}[1]{{{#1}}}
\newcommand{\gcc}[1]{{{#1}}} 
\newcommand{\newshang}[1]{{{#1}}} 

\newcommand{\fanyii}[1]{{{#1}}}

\def\BibTeX{{\rm B\kern-.05em{\sc i\kern-.025em b}\kern-.08em
    T\kern-.1667em\lower.7ex\hbox{E}\kern-.125emX}}
\begin{document}

\title{\shang{Towards practical and massively parallel quantum computing emulation for quantum chemistry} \\
}

 \author{\IEEEauthorblockN{Honghui Shang\IEEEauthorrefmark{2}\IEEEauthorrefmark{6}\IEEEauthorrefmark{1}, 	
         Yi Fan  \IEEEauthorrefmark{3}\IEEEauthorrefmark{6},
 		Li Shen\IEEEauthorrefmark{3},
 		Chu Guo  \IEEEauthorrefmark{5}\IEEEauthorrefmark{1},
 		Jie Liu  \IEEEauthorrefmark{3}\IEEEauthorrefmark{1},
 		Xiaohui Duan\IEEEauthorrefmark{4}, 
 		Fang Li \IEEEauthorrefmark{8}, 
 		Zhenyu Li\IEEEauthorrefmark{3}
 		}
 	\IEEEauthorblockA{\IEEEauthorrefmark{2} Institute of Computing Technology, Chinese Academy of Sciences, Beijing, China}
 	\IEEEauthorblockA{\IEEEauthorrefmark{3}Hefei National Laboratory, University of Science and Technology of China, Hefei 230088, China}

  \IEEEauthorblockA{\IEEEauthorrefmark{5}Key Laboratory of Low-Dimensional Quantum Structures and Quantum Control of Ministry of Education,}
  \IEEEauthorblockA{Department of Physics and Synergetic Innovation Center for Quantum Effects and Applications,}
  \IEEEauthorblockA{Hunan Normal University, Changsha 410081, China}

  \IEEEauthorblockA{\IEEEauthorrefmark{4}School of Software, Shandong University, Jinan, China}
  
  \IEEEauthorblockA{\IEEEauthorrefmark{8}National Research Center of Parallel Computer Engineering and Technology, Beijing, China}

 	\IEEEauthorblockA{\IEEEauthorrefmark{6}These authors contribute equally to this work; \IEEEauthorrefmark{1}Corresponding authors}
 }


\maketitle
\thispagestyle{plain}
\pagestyle{plain}

\begin{abstract}
Quantum computing is moving beyond its early stage and seeking for commercial applications in chemical and biomedical sciences. In the current noisy intermediate-scale quantum computing era, quantum resource is too scarce to support these explorations. Therefore, it is valuable to emulate quantum computing on classical computers for developing quantum algorithms and validating quantum hardware. However, existing simulators mostly suffer from the memory bottleneck \newshang{so developing the approaches for large-scale quantum chemistry calculations remains challenging.}
 Here we demonstrate a high-performance and \fanyii{massively parallel} variational quantum eigensolver (VQE) simulator based on matrix product states, combined with embedding theory for solving \newshang{large-scale quantum computing emulation for quantum chemistry} on HPC platforms. We apply this method to study the torsional barrier of ethane and the quantification of the protein-ligand interactions. Our largest simulation reaches $1000$ qubits, and a performance of $216.9$ \fanyii{PFLOPS} is achieved on a new Sunway supercomputer, which sets the state-of-the-art for \newshang{quantum computing emulation for quantum chemistry}.

\end{abstract}


%
%



\section{Introduction}
Computation is revolutionizing chemistry and materials science. Computing the electronic structure by approximately solving the Schr\"odinger equation enables us to explore chemicals and materials at the atomic scale. However, the pursuit for chemical accuracy in numerical simulations of quantum many-body systems is a longstanding problem since the computational complexity grows exponentially with the system size. For example, even with the help of supercomputers, the exact solution of the Schr\"odinger equation is limited to a complete active space problem of ($24$ electrons, $24$ orbitals), which corresponds to a diagonalization problem of size $7.3$ trillion~\cite{VogMaOls17}. Richard Feynman suggested quantum computing as a potential solution for simulating quantum systems, as he marked \fanyii{`if you want to make a simulation of nature, you’d better make it quantum mechanical'}~\cite{feynman2018simulating}. 

Significant advances in quantum computing technologies over the past two decades are turning Feynman's vision into reality. As a milestone, quantum advantage in the random circuit sampling (RCS) problem has been demonstrated on noisy intermediate-scale quantum (NISQ) computers~\cite{AruteMartinis2019,WuPan2021,ZhuPan2022}. Toward practical applications, the ground-state energies of diamond have been estimated with the quantum Monte Carlo (QMC) method using 16 qubits and 65 circuit depths, which is the largest \newshang{quantum chemistry calculation} using a quantum computer~\cite{Huggins2022}. However, the quantum resource used in this experiment is far away from that required to realize quantum advantage in quantum chemistry, which is expected to appear at around $38$ to $68$ qubits (under the assumption of error-corrected qubits)\cite{ElfBroWeb20}. Besides, the variational quantum eigensolver (VQE) is an appealing candidate for solving  \newshang{quantum chemistry} problems on NISQ devices~\cite{PerMcCSha14}, which has a great flexibility in choosing quantum circuit ansatzes and mitigating errors~\cite{McardleYuan2020}. However, compared to the RCS and QMC experiments, the VQE simulations with tens of qubits would be significantly more challenging for quantum hardware in that: 1) the circuit depth scales quickly up to $10^3$ or even more as the number of qubits increases~\cite{Michael_Resource_est_2019}; and 2) the nonlinear optimization with a large number of parameters remarkably increases the computational cost. As such, the largest VQE experiment preformed on a quantum computer has only used 12 qubits~\cite{Google2020a}, and the current VQE emulation with classical simulators are also mostly limited to relatively small molecules with 10 to 20 qubits, as shown in Table~\ref{table:overview} for the typical simulations of chemical and material systems using classical simulators.

To explore practical applications of quantum computing in quantum chemistry, one can resort to the development of quantum technologies, e.g. advanced quantum algorithms in combination with error mitigation techniques or the fault-tolerant quantum computers as a long-term target. Another way is the combination of state-of-the-art simulators with high performance computing (HPC), which enable us to emulate large-scale quantum computation of the electronic structure on classical computers. In the current stage, simulators are expected to play a fundamental role in the algorithm design or verification. In the RSC experiments, classical simulators are used for both calibrating the fidelity of individual gate operation and the whole random quantum circuit, and extrapolating the fidelity of simpler quantum circuits to the most difficult ones~\cite{AruteMartinis2019,WuPan2021,ZhuPan2022}. In most quantum algorithm designs, simulators are employed as the numerical emulating platform to benchmark new algorithms. 

Classical simulators suffer from the notorious exponential wall when the many-body systems are simulated exactly. As such, approximation algorithms are often used to realize large-scale emulations of \newshang{quantum chemistry calculation}. \liu{For example, the excited states of iridium complexes have been computed with up to $72$ qubits~\cite{GenRyaPai22}, which is the largest classical emulation of the VQE in terms of the number of qubits up to date. However, to achieve such a large emulation scale, a very shallow quantum circuit ansatz was employed to reduce the computational cost.} Additionally, a $28$-qubit VQE emulation of the \fanyii{C$_2$H$_4$} molecule has been reported by using point symmetry to significantly reduce the total number of gate operations~\cite{CaoYung2022}. \liu{A classical emulation of the \fanyii{C$_{18}$} molecule (a model system consisting of 144 spin molecular orbitals and 72 electrons) has been reported by combining VQE with the density matrix embedding theory (DMET), where DMET is used to break the molecule into small fragments and the VQE is used as the solver for the electronic structure of each fragment. While, the maximum number of qubits used in the VQE calculations is only 16~\cite{LiLv2022}.}


In this work, we demonstrate a high-performance and \fanyii{massively parallel} VQE simulator using the matrix product state (MPS) representation of the quantum state. Our simulator maximally utilizes the power of tensor network methods and supercomputers in order to to overcome the exponential memory bottleneck and realize the largest classical emulation of quantum computational chemistry.  The major computational bottleneck of the MPS-VQE algorithm \shang{(see Sec.~\ref{ss:mps-algo} for more details)} on HPC is the implementation of high-level linear algebra solvers, such as singular value decomposition (SVD)~\shang{(see Sec.~\ref{ss:SVD})}. Here, we overcome this bottleneck by \fanyii{the optimized} SVD and tensor operation algorithm. As discussed in Sec.~\ref{subsection:speedup}, our one-sided Jacobi SVD is more than $60$ times faster than the non-optimized version on average for matrix sizes from $100$ to $500$. As a result, our largest simulation which use the MPS-VQE simulator scales up to $1000$ qubits for one-shot energy evaluation and to $92$ qubits for \fanyii{fully} converged VQE emulation, with a two-qubit gate count up to $10^5$. In combination with DMET~\shang{(see Sec.~\ref{ss:DMET} for more details)}, our simulator is applied to study practical quantum chemistry systems containing $103$ atoms and achieves comparable accuracy with state-of-the-art computational methods.




\section{Results}
\subsection{Optimization Strategies}
\label{subscton:many-core}
Emulating quantum computing on a classical computer is difficult due to the exponential runtime and memory requirement. Such \fanyii{difficulties} can be leveraged with tensor network methods and by utilizing many-core and multi-node computers. Heterogeneous many-core systems are efficient for handling runtime issues but have limited total accessible memory space. Meanwhile, the memory of a multi-node computer can be scaled to the petabytes order, but its bandwidth for access from host computers (CPUs) is narrow. To simultaneously accelerate simulations and enlarge the total memory space, the heterogeneous parallelization approach~\cite{ShangShen2022}(see Sec.~\ref{ss:parellel} and Sec.~\ref{ss:julia} for more details) can be adopted. Our simulator allocates memory to each \fanyii{computation} node and then accelerates simulations by utilizing the full capabilities of the heterogeneous many-core processors. 

The new-generation Sunway supercomputer that is the successor of the Sunway TaihuLight supercomputer is used for performance assessment in this work. Similar to the Sunway TaihuLight system, the new Sunway supercomputer adopts a new generation of domestic  high-performance heterogeneous many-core processors~(SW26010Pro) and interconnection network chips in China. The architecture of the SW26010Pro processor is shown in Fig.~\ref{fig:sw-zgemm-svd}(a). Each processor contains 6 core-groups (CGs), with 65 cores in each CG, making a total \fanyii{number} of 390 cores. Each CG contains one management processing element (MPE), one cluster of computing processing elements (CPEs) and one memory controller. Each CPE has a 32 KB L1 instruction cache, and a 256 KB  scratch pad memory (SPM, also called the Local Data Memory (LDM)), which serves the same function as the L1 cache. Data transfer between LDM and main memory can be realized by direct memory access (DMA).


The hotspots of our simulator are mainly the tensor contractions and SVD functions.  
In the tensor contraction, the first step is the index permutation of the tensors, \shang{followed by one of the BLAS~(basic linear algebra subprograms)~\cite{Blackford2002} routine that performs matrix-matrix multiplications~(ZGEMM) to accomplish the calculation}. Here we use the fused permutation and multiplication technique\cite{Liu2021}. For the ZGEMM calculation, we perform matrix-matrix multiplications based on the optimization strategies, including balanced block \shang{that we choose an optimized block for the matrix A and B to make balanced computations with CPEs}, and diagonal broadcasting method \shang{where we use CPEs on the diagonal to perform a broadcast to forward its data to its corresponding row or column}, to realize efficient parallel computing for matrix multiplications, matrix transpose multiplications and conjugate transpose multiplications on the Sunway many-core system. First, we need to decompose the matrices A and B into smaller blocks to fit into the computing size of kernel. Second, we transmit the blocks of the input matrix into the LDM from the main memory. If we need to permute the input matrix, we should load the data that need to be transposed to the LDM of each CPE in blocks by DMA\_get, and the data stored on its own LDM using the SIMD~\shang{(Single Instruction Multiple Data)} \fanyii{`vshuff'} instruction~\shang{(the interface of the shuffle between two vectors)}; A diagonal broadcast optimization method is used to greatly reduce the memory access overhead to ensure the overall performance of matrix multiplication. Third, SIMD is used to implement eight 64-bit double-precision floating-point operations at a time. One SIMD instruction is equivalent to a small loop, so the number of instructions can be reduced, thereby reducing the requirement for bandwidth, and reducing the number of loops caused by induced control-related time overhead, as shown in Fig.~\ref{fig:sw-zgemm-svd}(b).


\shang{ 
For the SVD calculation, there are mainly two classes of algorithms. The first class of the SVD algorithms is the QR-based two-phase approach\cite{gu1994efficient}, in which the matrix $A$ is transformed into a bidiagonal matrix using an orthogonal transformation, and then the bidiagonal matrix is diagonalized using the bidiagonal divide-and-conquer method or the QR algorithm. The complete SVD is then determined during the backward transformation. This method is efficient for large matrices while suffering from loss of relative accuracy~\cite{Demmel1992}. The second class of the SVD algorithms is the Jacobi-based algorithm, which has recently attracted a lot of attention because it has a higher degree of potential parallelism~\cite{Martin2015,Novakovic2011,Lahabar2009}. There are two varieties of the Jacobi-based algorithm~(see Sec.~\ref{ss:SVD}), one-sided and two-sided algorithms. The one-sided Jacobi algorithm is computationally more efficient than the two-sided algorithm~\cite{Dongarra2018}, and suitable for vector pipeline computing. Thus, to achieve efficient parallel SVD computation on Sunway heterogeneous many-core architectures, the best choice is the Hestenes one-sided Jocobi transformation method~\cite{Hastens1958}, where all pairs of columns are repeatedly orthogonalized in sweeps using Jacobi rotations~\cite{doi:10.1137/0910023} until all columns are mutually orthogonal. When the convergence is reached, the right singular vectors can be computed by accumulating the rotations, the left singular vectors are the normalized columns of the modified matrix, and the singular values are the norms of those columns. Since each pair of columns can be orthogonalized independently, the method is also easily to parallelize over the CPEs, as shown in \newshang{Fig.~\ref{fig:sw-zgemm-svd}(c)}. 
It should be noted that another scalable SVD algorithm called cross-product SVD~\cite{SchDrew20svd} is also widely used in principal component analysis. However, numerical issues may appear since the condition number is squared in the intermediate step to orthogonalize $A^T A$. To simulate quantum systems in which the superposition of states is quite arbitrary, the cross-product SVD may be not as stable as other approaches.}



\subsection{Validation Results with MPS-VQE simulator~(92 qubits)}
As a pilot application, Fig.~\ref{fig:h2} shows the potential energy curves (PECs) of the hydrogen molecule computed with the MPS-VQE simulator. The unitary coupled cluster with single and double excitations (UCCSD) ansatz that is able to accurately describe this two-electron system is employed for single-point energy calculations. \liu{The implementation of the UCCSD ansatz with MPS is described in Method (see Sec.~\ref{ss:ansatz_ucc} for more details).} The STO-3g, cc-pVDZ, cc-pVTZ and aug-cc-pVTZ basis sets are used to extend these emulations from 4 to 92 qubits. \fanyi{The BOBYQA optimizer is used for the variational optimization, with a convergence threshold set to $10^{-6}$ for the minimum allowed value of trust region radius.} \liu{Note that the hydrogen molecule can be simulated without supercomputer resources even in aug-cc-pVTZ basis, since only two electrons are involved. \fanyii{However, this 92-qubit case involves $1.4\times 10^5$ CNOT gates (161 variational parameters), which is the largest quantum circuit simulation up to date in terms of the number of qubits and circuit depth.}}  \newshang{The simulations are carried out using 512 processes, and the computation times are given in Tab.~\ref{table:h2-timing}.} 
\fanyi{The results from MPS-VQE are in excellent agreement with the full configuration interaction (FCI) results as shown in Table~\ref{table:conv}. For all the four basis sets, chemical accuracy is achieved with a maximum error of 0.82 \fanyii{kcal mol$^{-1}$} at R(H-H)=2.4 \r{A} for the aug-cc-pVTZ results.}
We also show results obtained with FCI in the complete basis set (CBS) limit, which can be considered as the exact potential energy curve of the hydrogen molecule. \fanyi{The results of aug-cc-pVTZ shows an average deviation of 1.42 \fanyii{kcal mol$^{-1}$} from the complete basis set limit.} We can see that using a larger basis set makes the potential energy curve much closer to the exact dissociation limit. 






\subsection{Speedup and Scaling with MPS-VQE simulator}
\label{subsection:speedup}

One major bottleneck of the MPS-VQE simulator is the SVD function (technical details shown in Sec.\ref{ss:SVD}), which takes around 85\% of the CPU time on average. 
\shang{In Fig.~\ref{fig:scaling}, we show the performance improvement of the two optimized versions of SVD, including the QR-based method implemented in SW\_xmath (QR\_SW\_xmath) and the optimized one-sided Jacobi in this work (one-sided-Jacobi\_SW), compared to the QR-based SVD method running on MPE (QR\_MPE), for different matrix sizes. We use the performance of the QR\_MPE as the baseline, which we set as 1 in Fig 4(b). We can see that the optimized SVD using the one-sided Jacobi method produces an overall speedup ranging from 1.5× to 62.2× compared to QR\_MPE, and achieves a speedup of 2x to 6× compared to QR\_SW\_xmath version. For the one-sided Jacobi SVD (one-sided-Jacobi\_SW), we use the Athread library routines provided by the Sunway architecture for the many-core acceleration, and we use 64 threads for the actual computation.  The Jacobi-based method for SVD used in this work has potentially better accuracy than other methods. For example, if the SVD routine in the MPS simulator is replaced with cross-product SVD\cite{SchDrew20svd}, the energy error with respect to FCI will raise from $1.1\times10^{-2}$ \fanyii{kcal mol$^{-1}$} to $1.5\times10^{-1}$ \fanyii{kcal mol$^{-1}$} for the simplest H$_2$ molecule (cc-PVTZ basis set) even if more than 2.5 times the number of VQE steps are performed. 

For the tensor contraction using the optimization method listed in Sec.~\ref{subscton:many-core}~(SW\_zgemm), we can get an overall speedup of around 1.3× to 7.2x compared with the SW\_xmath version~\shang{(a vendor-provided linear algebra library on the Sunway \fanyi{supercomputer})}, as shown in Fig.~\ref{fig:scaling}(a). 
}


Figure~\ref{fig:scaling}(c) shows the computational time of the MPS-VQE simulator for implementing the VQE circuits of the hydrogen chain using 512 processes. The maximally allowed bond dimension is set to be $D=128$, \fanyi{as explained in Section~\ref{ss:ansatz_adapt}}. \shang{The one-shot energy estimation means that only one step of energy evaluation is performed instead of performing optimization of variational parameters until convergence. In the one-shot energy evaluation, the parameters are set as random numbers in order to keep the bond dimension at the upper limit value (D=128) during the circuit evolution. }The number of the electrons/atoms ranges from 12 to 500, and the corresponding number of the qubits ranges from 24 to 1000. The scaling exponents of the computation time (as a function of the total number of atoms $N$) for each VQE iteration are fitted by the polynomial scaling formula $t=cN^{\alpha}$ ($\alpha$ is the exponent). We find the exponent $\alpha \approx 1.6$ for all of the VQE circuits. This is because the number of terms in the Hamiltonian approximately scales as $N^{1.5}$ for the hydrogen chain. 


\subsection{Peak performance with DMET-MPS-VQE}

We use the hydrogen chain to assess the scalability and performance of our DMET-MPS-VQE simulator. \liu{ The wave function ansatz is adaptively built in order to reduce the circuit depth (see Sec.~\ref{ss:ansatz_adapt} for more details). The system is divided into fragments with the DMET method. A brief introduction of the DMET method used in this work can be found in Sec.~\ref{ss:DMET}.} We record the computational time with an increasing number of fragments (2048 processes per fragment). 
The number of floating point operations for tensor contractions is measured by counting all the floating point arithmetic instructions needed for matrix multiplications. For SVD, the number of floating point operations is measured using the profiler LWPF~\cite{Duan2020} that can monitor the floating-point operation hardware counters in the processor. The quantum circuits containing CNOT gates acting on each pair of neighbouring qubits. This building block serves as the entanglement blocks in the hardware-efficient ansatz~\cite{model-circ-ref}. Evolving the circuit requires to perform SVDs for $N_{q}-3$ matrices of size $2D\times 2D$ and $3\times(N_{q}-3)$ matrix-matrix multiplications. The results are shown in Fig.~\ref{fig:scaling}(d). We can see that a nearly linear scaling is obtained. Sustained performance of 216.9 \fanyii{PFLOPS} is achieved in double precision with 606,208 processes~(39,403,520 cores) for the system with 2368 qubit. 

\subsection{Implications}
In this section, we discuss applications of our MPS-VQE and DMET-MPS-VQE simulators to study realistic chemical systems. One example is the torsional barrier of ethane, which is one of the most fundamental problems in biomacromolecule configuration analysis. Fig.~\ref{fig:ethane} shows the results obtained by the MPS-VQE simulator for the torsional barrier of the ethane molecule. The bond lengths of C-C and C-H are set to be 1.512 and 1.153 \r{A}, respectively. The STO-3g basis set with all 16 orbitals is used~(32 qubits). The obtained torsional barrier is \fanyi{0.29 eV which is} higher than the experimental value \fanyi{0.13 eV}. \fanyi{Using the 6-31G(d) basis set will lower the barrier to 0.20 eV even if a small active space of only 6-orbital-6-electron is used. Therefore, It is expected that using a larger basis set could further improve the simulation accuracy.}

As an anticipated application, we apply the DMET-MPS-VQE simulator to study the quantification of the protein-ligand interactions, which is a large-scale practical biochemical problem. Compared to the classical calculations, the quantum mechanical calculations can automatically include the effects of polarization, charge transfer, charge penetration, and the coupling of the various terms, thus offering more accurate and detailed information on the nature of the protein-ligand interactions. This is highly important in high-accuracy binding affinity prediction as well as in drug design. \shang{The SARS-CoV-2 is the  coronavirus behind the COVID-19 pandemic, and its main protease~(M$^{\rm pro}$) is an enzyme that cleaves the viral polyproteins into individual proteins required for viral replication,} so it is important to develop drugs targeting at M$^{\rm pro}$ for SARS-CoV-2. 
In quantum mechanical studies, the protein-ligand binding energy is calculated by $E_{\rm b}=E_{\rm complex}-E_{\rm protein}-E_{\rm ligand}$, where $E_{\rm complex}$ is the energy of the complex,  $E_{\rm protein}$ is the energy of the protein 
and $E_{\rm ligand}$ is the energy of unbound ligand.  
The energy of the complex, protein and the ligand bounded in the complex are calculated using density functional theory with the PBE+MDB functional to account for many-body van der Waals interactions, which is important to obtain accurate potential-energy surfaces~\cite{HojaTkatchen2017}. After that, the energy differences between bounded and unbounded geometries of ligands are estimated with DMET-VQE~\cite{Kirsopp2021}.
We use the geometries of the 14 neutral ligands from Ref.~\citenum{Wang2021}, and then we optimize the geometries of the ligands at the Hartree-Fock level to account for geometric distortion needed for the ligand to occupy the active site. \shang{Similar with Ref.~\cite{Kirsopp2021}, we use STO-3g basis set in the \fanyii{DMET-VQE calculation}.} We plot the ranking score against the experimental binding free energies in a correlational plot as shown in Fig.~\ref{fig:scaling}. The ranking score is defined as the difference between the binding energy and the average value of 14 ligands. Ideally, the simulated ranking score should reproduce the experimental trends. We use the coefficients of determination, denoted as R$^2$, of the simulated ranking score and the experimentally measured free energy to access the quality of our simulation. It can be seen the correlation between our simulation and the experiment is fairly good, with R$^2$ of 0.44, which is better than the FEP-based approach (with R$^2$ of 0.29)~\cite{LiZhe2020}. The dipyridamole falls off the correlation line, but the fact that candesartan cilexetil binds best to the protein agrees with experiment. By removing dipyridamole and hydroxychloroquine from the set, we get an R$^2$ of 0.59.  \liu{However, we are fully aware of the necessity to consider the basis set, environment and temperature effects, as well as DMET subsystems size when applying the DMET-MPS-VQE to drug design in the following studies.} The largest molecule we calculated is Atazanavir which contains 103 atoms and 378 electrons, this is the largest system that has been investigated with simulators to our knowledge.

\section{Discussions}

As a heuristic quantum algorithm, the accuracy and performance of VQE should be verified in practical applications. The problems that VQE aims to solve, namely finding the ground state of a quantum many-body Hamiltonian, have a computational complexity growing exponentially with the problem size in general. Therefore, small scale simulations for simple molecules using around $20$ qubits are hard to demonstrate the powerfulness of VQE in practical applications.
In this work, the MPS-VQE simulator scales up to $1000$ qubits for one-shot energy evaluation and to $92$ qubits for converged VQE emulation, moreover, the DMET-MPS-VQE simulator scales up to $39$ million cores on the New Sunway supercomputer. 
The quantification of the protein-ligand interactions for SARS-CoV-2 is studied with the \fanyii{DMET-MPS-VQE} as an application in drug discovery. Particularly, we can obtain descent results using VQE, which are comparable with the experimental observations.


The development of quantum computers requires the intertwining and contribution from classical supercomputers, which enables us to benefit from the much more mature classical computing.
The simulation scale we have reached in this work, in terms of both the number of qubits and the circuit depths, is far beyond the simulations that have done in existing literatures, and the capability of existing quantum computers. Although we have limited ourselves to the physical motivated UCCSD  ansatz, our simulator could also be straightforwardly used with any other circuit ansatz, such as those hardware efficient ones, which are more friendly to current quantum computers. Our simulator would be an excellent benchmark and validation tool for the development of next generation quantum computers, as well as a flexible platform for quantum researchers to explore industrially related applications with tens of qubits.

\section{Methods}
\liu{
\subsection{Unitary coupled cluster}
\label{ss:UCC}
The electronic Hamiltonian $\hat{H}$ of a chemical system is written in the second-quantized form as $\hat{H}=\sum_{p,q}{h_{q}^{p}a_{p}^{\dagger}a_{q}} +\frac{1}{2}{\sum_{\substack{p,q\\r,s}}}{g_{rs}^{pq}a_{p}^{\dagger}a_{q}^{\dagger}a_{r}a_{s}}$, where $h_{q}^{p}$ and $g_{rs}^{pq}$ are one- and two-electron integrals in the Hartree-Fock orbital basis.
In the framework of the VQE, the total energy is calculated by measuring the expectation values of the qubit Hamiltonian obtained by Fermion-to-Qubit transformations, such as Jordan-Wigner or Bravyi-Kitaev, of the fermionic Hamiltonian.
One of the most widely used wave function ansatz is the unitary \fanyii{coupled cluster}~\cite{McardleYuan2020} in the form of $|\Psi(\theta)\rangle=e^{\hat{T}(\theta)-\hat{T}^{\dagger}(\theta)}|\Phi_{0}\rangle$. Here, $|\Phi_{0}\rangle$ is the Hartree-Fock state, which can be easily prepared on a quantum computer. When the UCC operator is truncated to the single and double excitations (UCCSD), namely
\begin{align}
    \hat{T}(\theta)&=\sum_{a,i}\theta_{i}^{a}\hat{a}^\dag_{a} \hat{a}_{i} +\frac{1}{4}{\sum_{a,b,i,j}} \theta_{i j}^{a b} \hat{a}_{a}^{\dagger}\hat{a}_{b}^{\dagger} \hat{a}_{i} \hat{a}_{j},
\label{eq:ucc}
\end{align}
where $\{i, j, \dots \}$, $\{a, b, \dots\}$ and $\{p, q, \dots\}$ denote the occupied, virtual, and general spin molecular orbitals, respectively. The UCCSD ansatz does not have an exact finite truncation of the Baker-Campbell-Hausdorff expansion such that an approximation should be introduced in its classical implementation. 

The UCCSD ansatz can be implemented on a quantum platform with a parametric quantum circuit generated from Suzuki-Trotter decomposition of the unitary exponential operator into one- and two-qubit gates~\cite{PouHasWec15}. In such a case, the UCCSD ansatz can be mapped to a W-shaped ansatz circuit with a quartic number of two-qubit gate. For example, restricting to the minimal basis set, the number of CNOT gate of a full UCC circuit reaches $8.6\times10^5$ for the simple C$_2$H$_4$ molecule, which is usually far beyond the capability of current NISQ devices and can hardly be simulated on most of existing quantum circuit simulators.

We note that UCCSD is inadequate for describing many strongly correlated systems. Here, we focus on exhibiting the performance of our simulator. The accuracy of the wave function ansatzes can be improved by introducing adaptive VQE algorithms~\cite{GrimsleyMayhall2019}. 
}

\subsection{MPS algorithm for quantum circuit simulation}\label{ss:mps-algo}

The correlated wave function in quantum chemistry considering all configuration states can be written as:

\begin{align}
\left|\Psi\right\rangle  = \sum_{i_1...i_N} c_{i_1 i_2 i_3 \ldots i_N} |i_1 i_2 i_3 \ldots i_N \rangle
\end{align} 
where $|i_1 i_2 i_3 \ldots i_N \rangle$ refers to the computation basis, $c_{i_1 i_2 i_3 \ldots i_N}$ is a rank-N tensor of $2^N$ complex numbers. This state can be represented with matrix product states (MPS), decompose the correlated wave function into a set of low-rank tensors:
\begin{align}\label{eq:mps}
    c_{i_1 i_2 i_3 \ldots i_N} = \sum_{\alpha_0...\alpha_{N}}  
    B^{i_1}_{\alpha_0 \alpha_1} B^{i_2}_{\alpha_1\alpha_2}
    B^{i_3}_{\alpha_2\alpha_3} \ldots  B^{i_{N}}_{\alpha_{N-1}\alpha_N},
\end{align}
where $i_n \in \{0,1\}$ refers to ``physical'' indices and $\alpha_n$ the ``virtual'' index related to the partition entanglement entropy. $\alpha_0$ and $\alpha_N$ at the boundaries are trivial indices added for notational convenience. 


In our MPS simulator, we keep the tensors to be right-canonical, \gcc{namely the site tensors of the MPS in Eq.(\ref{eq:mps}) satisfy}:
\begin{align}\label{eq:rightcanonicalcondition}
\sum_{i_n, \alpha_n} (B^{i_n}_{\alpha_{n-1}', \alpha_n})^{\ast} B^{i_n}_{\alpha_{n-1}, \alpha_n} = \delta_{\alpha_{n-1}, \alpha_{n-1}'}.
\end{align}
A single-qubit gate operation acting on the $n$-th qubit, denoted as $Q_{i_n i'_n}$ can be simply applied onto the MPS as
\begin{align}
\tilde{B}^{i_n}_{\alpha_{n-1}\alpha_{n}}  = \sum_{i_n'} Q_{i_n i'_n} B^{i_n'}_{\alpha_{n-1}\alpha_{n}}.
\end{align}
The new site tensor $\tilde{B}^{i_n}_{\alpha_{n-1}\alpha_{n}}$ satisfies Eq.(\ref{eq:rightcanonicalcondition}) since $Q_{i_n i'_n}$ is unitary and $B^{i_n'}_{\alpha_{n-1}\alpha_{n}}$ satisfies Eq.(\ref{eq:rightcanonicalcondition}). 
For \gcc{the operation of} a two-qubit gate on qubits $n$ and $n+1$ (the $n$-th bond), denoted as $Q^{i_n,i_{n+1}}_{i'_n,i'_{n+1}}$, \gcc{we use the technique from Ref.~\cite{Hastings2009} to keep the underlying MPS in the right-canonical form, which is shown in the following.} We first contract the two site tensors $B^{i'_n}_{\alpha_{n-1},\alpha_n}$ and $B^{i'_{n+1}}_{\alpha_n,\alpha_{n+1}}$ with $Q^{i_n,i_{n+1}}_{i'_n,i'_{n+1}}$ to get a two-site tensor
\begin{align}
C_{\alpha_{n-1},\alpha_{n+1}}^{i_n,i_{n+1}}=\sum_{\alpha_n, i'_n,i'_{n+1}} Q^{i_n,i_{n+1}}_{i'_n,i'_{n+1}} B^{i'_n}_{\alpha_{n-1},\alpha_n} B^{i'_{n+1}}_{\alpha_n,\alpha_{n+1}},
\end{align}
then we contract $C_{\alpha_{n-1},\alpha_{n+1}}^{i_n,i_{n+1}}$ with the singular matrix formed by the singular values at the $n-1$-th bond (denoted as $\lambda_{\alpha_{n-1}}$) to get a new two-site tensor as
\begin{align}\label{eq:tildeC}
\tilde{C}_{\alpha_{n-1},\alpha_{n+1}}^{i_n,i_{n+1}}= \lambda_{\alpha_{n-1}} C_{\alpha_{n-1},\alpha_{n+1}}^{i_n,i_{n+1}}.
\end{align}
We perform singular value decomposition onto the tensor $\tilde{C}_{\alpha_{n-1},\alpha_{n+1}}^{i_n,i_{n+1}}$ and get
\begin{equation}\label{eq:svdtildeC}
\SVD(\tilde{C}_{\alpha_{n-1},\alpha_{n+1}}^{i_n,i_{n+1}})=\sum_{\alpha_n} U^{i_n}_{\alpha_{n-1},\alpha_n} \tilde{\lambda}_{\alpha_n} V^{i_{n+1}}_{\alpha_n,\alpha_{n+1}},
\end{equation}
during which we will also truncate the small singular values below a certain threshold or simply reserve the largest few singular values to control the memory overhead. Finally the new site tensors $\tilde{B}^{i_n}_{\alpha_{n-1},\alpha_n}$ and $\tilde{B}^{i_{n+1}}_{\alpha_n,\alpha_{n+1}}$ can be obtained as
\begin{align}
\tilde{B}^{i_n}_{\alpha_{n-1},\alpha_n} &= \sum_{i_{n+1},\alpha_{n+1}} C_{\alpha_{n-1},\alpha_{n+1}}^{i_n,i_{n+1}} \left(V^{i_{n+1}}_{\alpha_n,\alpha_{n+1}}\right)^{\ast}; \label{eq:Bl} \\
\tilde{B}^{i_{n+1}}_{\alpha_n,\alpha_{n+1}} &= V^{i_{n+1}}_{\alpha_n,\alpha_{n+1}},
\end{align}
and the new singular values $\tilde{\lambda}_{\alpha_n}$ is used to replace the old $\lambda_{\alpha_n}$ at the $n$-th bond. \gcc{Since $\sum_{\alpha_n} \tilde{B}^{i_n}_{\alpha_{n-1},\alpha_n} \tilde{B}^{i_{n+1}}_{\alpha_n,\alpha_{n+1}} = C_{\alpha_{n-1},\alpha_{n+1}}^{i_n,i_{n+1}} $, they indeed represent the correct site tensors after the two-qubit gate operation. $\tilde{B}^{i_{n+1}}_{\alpha_n,\alpha_{n+1}}$ is right-canonical by definition of SVD. Moreover, one can verify that $\tilde{B}^{i_n}_{\alpha_{n-1},\alpha_n}$ is also right-canonical by substituting Eqs.(\ref{eq:tildeC}, \ref{eq:svdtildeC}) into Eq.(\ref{eq:Bl}):
\begin{align}
\tilde{B}^{i_n}_{\alpha_{n-1},\alpha_n} = U^{i_n}_{\alpha_{n-1},\alpha_n} \tilde{\lambda}_{\alpha_n} / \tilde{\lambda}_{\alpha_{n-1}},
\end{align}
The above equation transforms a left-canonical site tensor $U^{i_n}_{\alpha_{n-1},\alpha_n}$ into a right-canonical site tensor $\tilde{B}^{i_n}_{\alpha_{n-1},\alpha_n}$.

}

\subsection{The implementation of UCCSD with matrix produce states}\label{ss:ansatz_ucc}
\liu{As discussed in Sec.~\ref{ss:UCC}, the implementation of the UCCSD ansatz in this work includes three step: 
\begin{itemize}
    \item We perform the Jordan-Wigner transformation of the cluster operator. Here, the Hartree-Fock state is employed as a reference state. The cluster operator is defined as a linear combination of single and double excitations from occupied orbitals to virtual orbitals (see Eq.~\eqref{eq:ucc}).  
    \item We perform a Suzuki-Trotter decomposition of the unitary exponential operator into one- and two-qubit gates. Because the excitation operators are not commutative, we use first-order Trotter decomposition to approximate the UCCSD ansatz as products of exponential operators, which can be further decomposed into products of one- and two-qubit gates. 
    \item We apply these quantum gates to a reference wave function. The intermediate wave functions after applying quantum gates to the initial wave function are represented by matrix product states. 
\end{itemize}
Step 1 and 2 are done using the Q$^2$Chemistry package\cite{q2chem_2022}. Step 3 is one of the most important parts of this work. Applying a single qubit gate to a MPS can be done without approximation by multiplying the gate with a single MPS tensor. To apply a two-qubit gate to qubits n and n + 1, we first perform tensor contractions the corresponding gates and tensors, and then applies the gate to the contracted state. To restore the MPS form, the resulting tensor is decomposed with a SVD truncated to keep the largest X singular values, and the matrix of singular values is multiplied into one of the unitary factors X or Y. 

With a right-canonical form of MPS, there is a very efficient way to compute the expectation of a single Pauli string. Taking the expectation value of a single-qubit observable $O_{i_n i'_n}$ as an example, it can be simply computed as
\begin{align}\label{eq:expec1}
\sum_{\alpha_{n-1},\alpha_n,i_n,i'_n} \lambda^2_{\alpha_{n-1}}O_{i_n i'_n} B^{i'_n}_{\alpha_{n-1},\alpha_n}(B^{i_n}_{\alpha_{n-1},\alpha_n})^{\ast},
\end{align}
while a generic two-qubit observable $O^{i_m, i_n}_{i'_m, i_n'}$ (assuming $m < n$) can be computed as
\begin{align}\label{eq:expec2}
\sum_{\alpha_{n:m-1},i_{n:m},i'_{n,m}} &\lambda^2_{\alpha_{m-1}}O_{i_m' i'_n}^{i_m i_n} B^{i'_m}_{\alpha_{m-1},\alpha_m}(B^{i_m}_{\alpha_{m-1},\alpha_m})^{\ast} \times \nonumber \\ 
&\dots \times B^{i'_n}_{\alpha_{n-1},\alpha_n}(B^{i_n}_{\alpha_{n-1},\alpha_n})^{\ast},
\end{align}
where we have used $x_{j:i} = \{x_i, x_{i+1}, \dots, x_j\}$ as an abbreviation for a list of indices.
The expectation value of a general $n$-qubit Pauli string could be computed similarly.
}

\subsection{The wave function ansatz for hydrogen chain simulations} \label{ss:ansatz_adapt}
\fanyi{When hydrogen chains containing hundreds of atoms are studied, it is impossible to implement a full UCCSD ansatz even with a supercomputer. As such, we construct approximate wave function ansatzes to perform such large-scale simulations using our simulator. The ansatzes are constructed following four steps.
\begin{itemize}
    \item The generalized single and double (GSD) excitation operators are generated using every 5 consecutive orbitals. For example, if there are 100 Hartree-Fock orbitals obtained from the Hartree-Fock calculation, we first build GSD excitation operators using orbital 1 to 5, and then orbital 2 to 6, etc.
    \item After the fermionic operator pool has been constructed, the Jordan-Wigner transformation is used to generate an initial operator pool $\{P\}$ in the form of Pauli strings.
    \item All the Pauli-Zs are removed from the Pauli strings in order to reduce the quantum circuit depth. Because the Hamiltonian is real, all Pauli strings with even number of Pauli-Ys are removed from $\{P\}$.
    \item The parametric circuit is adaptively constructed as a product of exponential of Pauli strings $\prod_{j} {\exp (\text{i} \theta_{j} P_{j})}$, where $P_{i}\in \{P\}$ and $\{\theta\}$ are variational parameters to be optimized. Here, we follows the strategy suggested in the qubit-ADAPT-VQE method~\cite{TangEconomou2021}. While we did not iteratively build the wave function ansatz until convergence, high accuracy can be achieved if more iterations are performed to improve the wave function ansatz.
\end{itemize}
The above steps are performed by interfacing our MPS-VQE simulator with the Q$^2$Chemistry package\cite{q2chem_2022,ShangShen2022}. In this way, an approximate wave function ansatz that entangles every neighbouring 5 orbitals (10 qubits) is constructed for the hydrogen chain simulations.}
\fanyi{Another important factor that affects the simulation accuracy is maximum allowed bond dimension of the MPS simulator. In order to choose a reasonable bond dimension, we performed a benchmark on the converged energy with respect to different bond dimension settings using a smaller molecule (H$_8$, 16 qubits). The results are given in Figure~\ref{fig:mps-d-benchmark} and the bond dimension is selected such that $\Delta E=|E_{D_i}-E_{D_{i+1} } |< 1.0\times10^{-3}$ Hartree which is slightly more strict than chemical accuracy ($1.6 \times 10^{-3}$ Hartree).}



\subsection{The DMET method}
\label{ss:DMET}
In DMET, a high-level calculation for each fragment~(e.g. VQE) is carried out individually until the self-consistency criterion has been met: the sum of the number of electrons of all of the fragments agrees with the number of electrons for the entire system. The DMET energy for the fragment is calculated using the 1-RDM and 2-RDM, that is,  
\begin{eqnarray}
	&E_{A} = \sum_{p \in A} \bigg(  \nonumber \\  &\sum_{q}^{N^A_{\text{orb}} + N^B_{\text{orb}}} \bigg(h_{pq} + \frac{\sum_{N_{\text{orb}}} [(pq|rs) - (ps|rq)] \Gamma^{\text{env}}_{rs}}{2}\bigg) D_{qp}^A  \nonumber \\ 
	&+ \frac12 \sum_{qrs}^{N^A_{\text{orb}} + N^B_{\text{orb}}} (pq|rs) P_{qp}^A \bigg),
\end{eqnarray}
where $h_{pq}$ are the one-electron integrals, $(pq|rs)$ are two-electron integrals, $N^A_{\text{orb}}$ is the number of orbitals in the fragment, $N^B_{\text{orb}}$ is the number of the bath orbitals, $N_{\text{orb}}$ is the total number of the orbitals in the entire molecule and p,q,r,s are orbital indices.  $D_{qp}^A=\langle \hat{a}^{\dagger}_p  \hat{a}_q\rangle$) is 1-RDM and and  $P_{qp}^A=\langle\hat{a}^\dagger_p \hat{a}^\dagger_q \hat{a}_r \hat{a}_s \rangle$ is 2-RDM, which are evaluated with VQE method in this work. The number of \fanyii{electrons} in fragment A is calculated as 
$N^{A}=\sum_{p A} D_{pp}^{A}$, and the DMET total energy is the sum of 
the fragment energies
\begin{equation}
	E^{\rm total}=\sum_A E_{A} 
\end{equation}
The DMET cycle iterates until the number of electrons $N^{\rm DMET}=\sum_A N^{A}$ converges to the total number of electrons in molecule ($N$) . 

\subsection{Heterogeneous parallelization strategy}
\label{ss:parellel}
For the DMET-MPS-VQE simulator, three levels of parallelization are adopted: (1) The calculation of different fragments can be performed in an embarrassingly parallel manner, that we split the whole CPU pool into different sub-groups and sub-communicators, and there is no communication between different fragment calculations;  (2) within each sub-group, the total energy of each fragment is calculated with the \fanyi{MPS-VQE} method.
\fanyi{We adopted the parallel simulation algorithm based on distributed memory over the circuits, just ``mimic'' the actual quantum computers, so our method can offer a good reference for VQE running on the quantum computers;} \fanyi{(3) within the simulations of a single quantum circuit,} we use a low-level multi-threaded parallelism on the CPEs to further boost the performance for the tensor contraction and singular value decomposition. We refer the reader to Ref.~\cite{ShangShen2022} for more details.

\subsection{Julia programming language}
\label{ss:julia}
The Julia script language is used as the main programming language in this study. Julia has the performance of a statically compiled language while providing interactive dynamic behavior and productivity~\cite{bezanson2012julia}. The codes written in Julia can be highly extensible due to its type system and the multiple dispatch mechanism. In additional to its JIT feature and meta-programming ability, its powerful foreign function interface (FFI) makes it easily to use external libraries written in other languages. In this study, the electronic structure libraries Pyscf\cite{pyscf} and OpenFermion\cite{openfermion} are linked to Julia through PyCall.jl, and the optimized SVD routines written in C is called using the LLVM.jl package which provides a high-level wrapper to the LLVM C API.

Our parallel algorithm implemented in Julia is based on the parallel libraries MPI.jl. MPI.jl is a basic Julia wrapper for the Message Passing Interface (MPI). On the Sunway architecture, the MPI libraries are versatile and are highly optimized. MPI.jl can call these MPI library through interfaces of Julia that are almost identical to the C language, and provides similar performance.

\subsection{SVD and one-side jacobi}
\label{ss:SVD}
The singular value decomposition of a 
Matrix $A_{m\times n}$ can be written as,
\begin{equation}
	A=U\Sigma V^T
\end{equation}
Where the matrix $A_{m\times n}$ is decomposed into three matrices. Matrix $U_{m\times m}$ and $V_{n\times n}$ are complex unitary matrices, and $V_{n\times n}^T$ is the conjugate transpose of $V_{n\times n}$. Matrix $\Sigma_{m\times n}$ is a rectangular diagonal matrix with the singular values of matrix $A_{m\times n}$ on the diagonal.

There are two classes of Jacobi-based SVD algorithms: one-sided and two-sided.  Two-sided Jacobi iteration algorithm transforms a symmetric matrix into a diagonal matrix by a sequence of two-sided Jacobi rotations ($J$).
\begin{equation}
	J(i, j, \theta)=\begin{blockarray}{cccccccc}
		\begin{block}{[ccccccc]c}
			1 & \cdots & 0 & \cdots & 0 & \cdots & 0 \\
			\vdots & \ddots & \vdots & & \vdots & & \vdots \\
			0 & \cdots & c & \cdots & -s & \cdots & 0 & i\\
			\vdots & & \vdots & \ddots & \vdots & & \vdots \\
			0 & \cdots & s & \cdots & c & \cdots & 0 & j\\
			\vdots & & \vdots & & \vdots & \ddots & \vdots \\
			0 & \cdots & 0 & \cdots & 0 & \cdots & 1 \\
		\end{block}
		& & i & & j 
	\end{blockarray}
\end{equation}
Based on two-sided Jacobi algorithm, one-sided Jacobi SVD calculates singular value decomposition with only one-sided Jacobi rotations that modifies columns only.   Algorithm \ref{algo:one-sided-jacobi-svd} describes the one-sided Jacobi method.
\begin{algorithm}
	\caption{One-sided Jacobi SVD method for $m \times n$ matrix $A$, $m \ge n$.}
	\label{algo:one-sided-jacobi-svd}
	
	\SetKwProg{Fn}{function}{}{}
	
	\Fn{\rm{one\_sided\_jacobi\_svd(A)}} {
		$V = I$ \BlankLine
		\For{i\_\rm sweep = 1, ..., N\_\rm sweep} {
			\BlankLine
			pass= true \BlankLine
			\For{\rm each column pair($i,j$), $i<j$}{
				$b_{ii} = A^\dagger_iA_i$ 
				\BlankLine
				$b_{jj} = A^\dagger_jA_j$ 
				\BlankLine
				$b_{ij} = A^\dagger_iA_j$ \BlankLine
				\If{$|b_{ij}| \ge \epsilon \sqrt{b_{ii}b_{jj}}$}{
					\BlankLine
					\rm get Jacobi rotation matrix $J$ \BlankLine
					$A = AJ$\\ 
				$V = VJ$ \BlankLine
					pass = false \BlankLine
				}
			}
			if(pass) break \BlankLine
		}
		\For{$i = 1,...,n$} {
			$\sigma_i = \sqrt{A_i^\dagger A_i}$  \BlankLine
			$u_i = a_i / \sigma_i$ \BlankLine
		}
		sort $\Sigma$ and apply same permutation to columns of $U$ and $V$ \BlankLine
		\Return{($U,\Sigma,V$)}
	}
\end{algorithm}
The parameters $c$ and $s$ of the Jacobi rotation matrix can be calculated by $t$ and $\tau$.
\begin{equation}    
	c=\frac{1}{\sqrt{1+t^2}}
\end{equation}
\begin{equation}
	s=t\times c
\end{equation}
\begin{equation}
	\ t=\frac{sign(\tau)}{\left|\tau\right|+\sqrt{1+\tau^2}}
\end{equation}
\begin{equation}
	\ \ \tau=\frac{a_i^Ta_i-a_j^Ta_j}{2a_i^Ta_j}
\end{equation}
The algorithm converges when all rotations in a sweep are skipped.
Since each pair of columns can be orthogonalized independently, the method is also easily parallelized over the CPEs. The simplicity and inherent parallelism of the method make it an attractive first choice for an implementation on the many-core system.

\subsection{The quantum simulation time of hydrogen chain with MPS-VQE}
\fanyii{The quantum simulation time of hydrogen chain using the MPS-VQE simulator is tested.} The number of atoms ($N_a$), number of qubits ($N_q$), and the estimated number of circuits ($N_c$) are listed in Tab.~\ref{tab:h2-line}. \newshang{The geometry of hydrogen molecule chain is set as following: the H$_2$ moieties with R(H-H) = 0.741 \AA ~was aligned, and the distance between closest atoms of different H$_2$ fragments was 1.322 \AA, as shown in Fig.~\ref{fig:h2-mol-chain}. For all the calculations, we use 512 cores (8 nodes~$\times$~64 cores per node).}



%

\section{DATA AVAILABILITY}
The data that support the findings of this study are available from the corresponding authors upon reasonable request.


\section{ACKNOWLEDGEMENTS}
H.S. acknowledges support from National Natural Science Foundation of China~(Grant No. T2222026, 22003073). L.S. acknowledges support from 
National Key Research and Development Program of China~(Grant No. 2018YFB0204200). C.G. acknowledges support from National Natural Science Foundation of China~(Grant No. 11805279).  J.L. acknowledges National Natural Science Foundation of China (Grant No. 22073086),  Innovation Program for Quantum Science and Technology (Grant No. 2021ZD0303306), the Fundamental Research Funds for the Central Universities (Grant No. WK2060000018). Computational  resources were provided by the new Sunway supercomputer. 

\section{COMPETING INTERESTS}
The authors declare no competing interests.

\section{AUTHOR CONTRIBUTIONS} 
The project was conceived by H.S. The manuscript was written by H.S., J.L. and C.G. The numerical simulations were performed by Y.F., H.S., L.S., F.L., X.D. and Z.L., H.S. and Y.F. contributed equally to this work and are considered as co-first authors.

\bibliographystyle{./IEEEtran}
\bibliography{refs.bib}

\input{tables.tex}
\input{figures.tex}

\end{document}

%% file: tables.tex
\begin{table}[!htb]
	\tabcolsep 1mm \caption{Typical simulations of molecular and material systems with classical simulators. Number of atoms ($N_a$), number of qubits ($N_q$), and the estimated number of CNOT gates ($N_{\rm CNOT}$) are listed for comparison. 
	}
	\begin{center}
		\begin{tabular}{cccccc}\hline \hline
			Work & System & $N_a$ & $N_q$ & $N_{\rm CNOT}$ &  Reference \\
			\hline
			Microsoft QDK  & H$_2$ & 2 & 4 & 696 & ~\citenum{bylaska2021quantum} \\ 
			Cirq  & CH$_2$O &  4 & 6 & $1.8\times 10^3$ & ~\citenum{YalSenGun21}\\ 
			Qulacs  &He crystal & 1 & 8 & $1.6\times10^3$ & ~\citenum{ManKhaYam21} \\ 
			Qiskit   &N$_2$ & 2 & 16 & $1.9\times10^4$ &~\citenum{XiaKai20} \\ 
			Yao.jl   & C$_{18}$ & 18 & 16 & $5.4\times10^4$ & ~\citenum{LiLv2022} \\ 
			VQEChem   & H chain & 2 & 16 & $5.4\times10^4$ &~\citenum{LiuWanLi20} \\ 
			QCQC  & Si crystal & 2 & 16 & $1.1\times10^5$ &~\citenum{FanLiuLi21} \\ 
			Tequlia  & BH & 2 & 22 & $6.2\times10^3$ &~\citenum{KotSchTam21} \\  
			HiQ  & C$_2$H$_4$ & 6 & 28 & $1.2\times10^5$ &~\citenum{CaoYung2022}\\ 
			iQCC-VQE  & Ir$^{\mathrm{III}}$ complexes & $\sim$60 & \liu{72} & \liu{$\sim 96$} &~\citenum{GenRyaPai22} \\ 
			\hline
			MPS-VQE &  H$_2$      & 2 &  92 & $1.4\times10^5$ &   \multirow{4}{*}{This work}        \\
			MPS-VQE &  C$_2$H$_6$      & 8 &  32 & $4.4\times10^5$ &     \\
			MPS-VQE~(one shot) &  H$_2$ chain     & 500 &  1000 & $1.0\times10^6$ &          \\ 
			DMET-MPS-VQE&  Atazanavier    & 103 & 16   & $1.8\times10^6$ & \\
			\hline \hline 
		\end{tabular}
	\end{center}
	\label{table:overview}
\end{table}

\begin{table}[!htb]
	\centering
	\caption{\newshang{Wall time per VQE iteration in seconds and number of iterations to converge for different basis sets. The data are collected from the geometry with the lowest energy of each basis set in Fig.~\ref{fig:h2}.}}
	\begin{tabular}{|c|cccc|}
		\hline
		\hline
		Basis set & STO-3g & cc-pVDZ & cc-pVTZ & aug-cc-pVTZ \\
		\hline
		Wall time per iteration & 0.12 & 3.67 & 190.63 & 1564.52 \\ 
		\hline
		Number of steps & 18 & 303 & 459 & 677 \\
		\hline
	\end{tabular}
	\label{table:h2-timing}
\end{table}

\begin{table}[!htb]
	\centering
	\caption{The mean absolute errors (MAE) and maximum absolute errors (MAX) (in kcal/mol) of the potential energy surfaces for H$_2$ compuated with the UCCSD-VQE method using different Gaussian basis sets. The FCI results are taken as the reference values.}
	\begin{tabular}{|c|cccc|}
		\hline
		\hline
		Basis set & STO-3g & cc-pVDZ & cc-pVTZ & aug-cc-pVTZ \\
		\hline
		MAE & 9.4$\times 10^{-13}$ & 2.7$\times 10^{-3}$ & 8.1 $\times 10^{-2}$ & 3.3 $\times 10^{-1}$ \\
		\hline
		MAX & 6.3$\times 10^{-12}$ & 1.3 $\times 10^{-2}$ & 1.8 $\times 10^{-1}$ & 8.2 $\times 10^{-1}$\\
		\hline
	\end{tabular}
	\label{table:conv}
\end{table}

\begin{table}
	\caption{The computational time per VQE iteration \newshang{using 512 cores} for the hydrogen chain with the MPS-VQE simulator (without DMET). The number of atoms ($N_a$), number of qubits ($N_q$), the estimated number of circuits ($N_c$) are listed in the  table. The bond dimension $D$ is set to be 128. }
	\centering
	\begin{tabular}{llllll}
		\hline \hline
		System  &  $N_a$  & $N_q$ & $N_c$ &   Wall time (s) & \newshang{CPU Time (core$\cdot$s)} \\
		\hline 
		(H$_2$)$_{3}$       & 6      & 12       & 1811     & 1.23 & 559.64              \\
		(H$_2$)$_{6}$       & 12     & 24       & 15905     & 4.20 & 2133.08             \\
		(H$_2$)$_{12}$      & 24     & 48       & 60723     & 10.67 & 5443.94            \\
		(H$_2$)$_{25}$      & 50     & 100      & 193607    & 29.90 & 14923.02            \\
		(H$_2$)$_{50}$      & 100    & 200      & 544549   & 86.74 & 43520.58            \\
		(H$_2$)$_{100}$     & 200    & 400      & 1426637  & 304.25 & 154234.77           \\
		(H$_2$)$_{250}$     & 500    & 1000     & 5059403   & 1961.03 & 999432.92 \\
		\hline \hline
	\end{tabular}
	\label{tab:h2-line}
\end{table}

%% file: figures.tex
\begin{figure*}
	\centering
	\includegraphics[width=1.9\columnwidth]{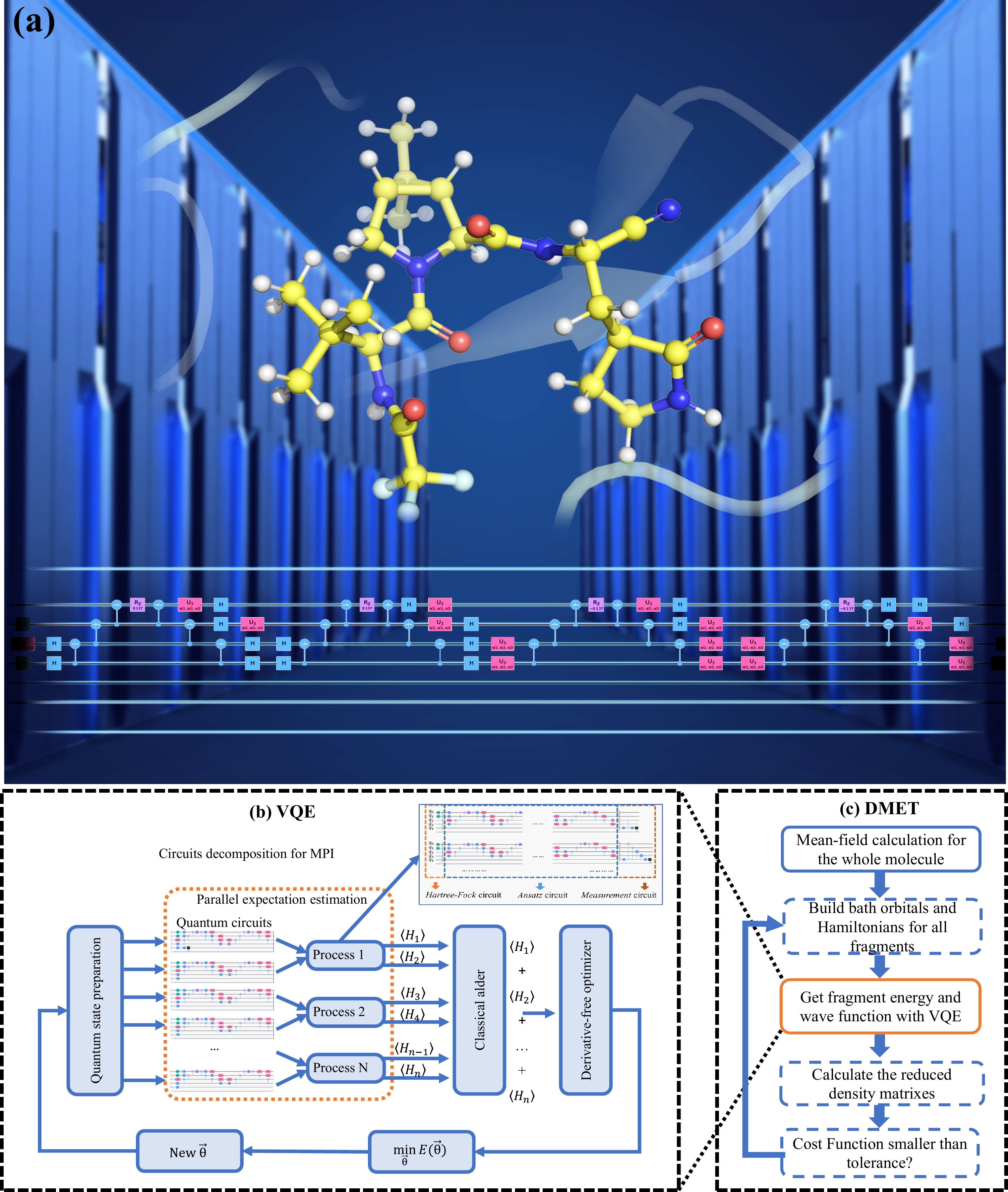}
	\caption{Framework of our Quantum Computational Chemistry
		simulator. (a) The conceptual illustration of the quantum computing emulation for quantum chemistry; (b) The VQE simulator using the matrix product states (MPS) representation of the quantum state for each fragment within DMET; (c) The DMET calculation procedures for the realistic chemical systems. }
	\label{fig:DFT-DMET-VQE}
\end{figure*}

\begin{figure*}
	\centering
	\includegraphics[width=1.98\columnwidth]{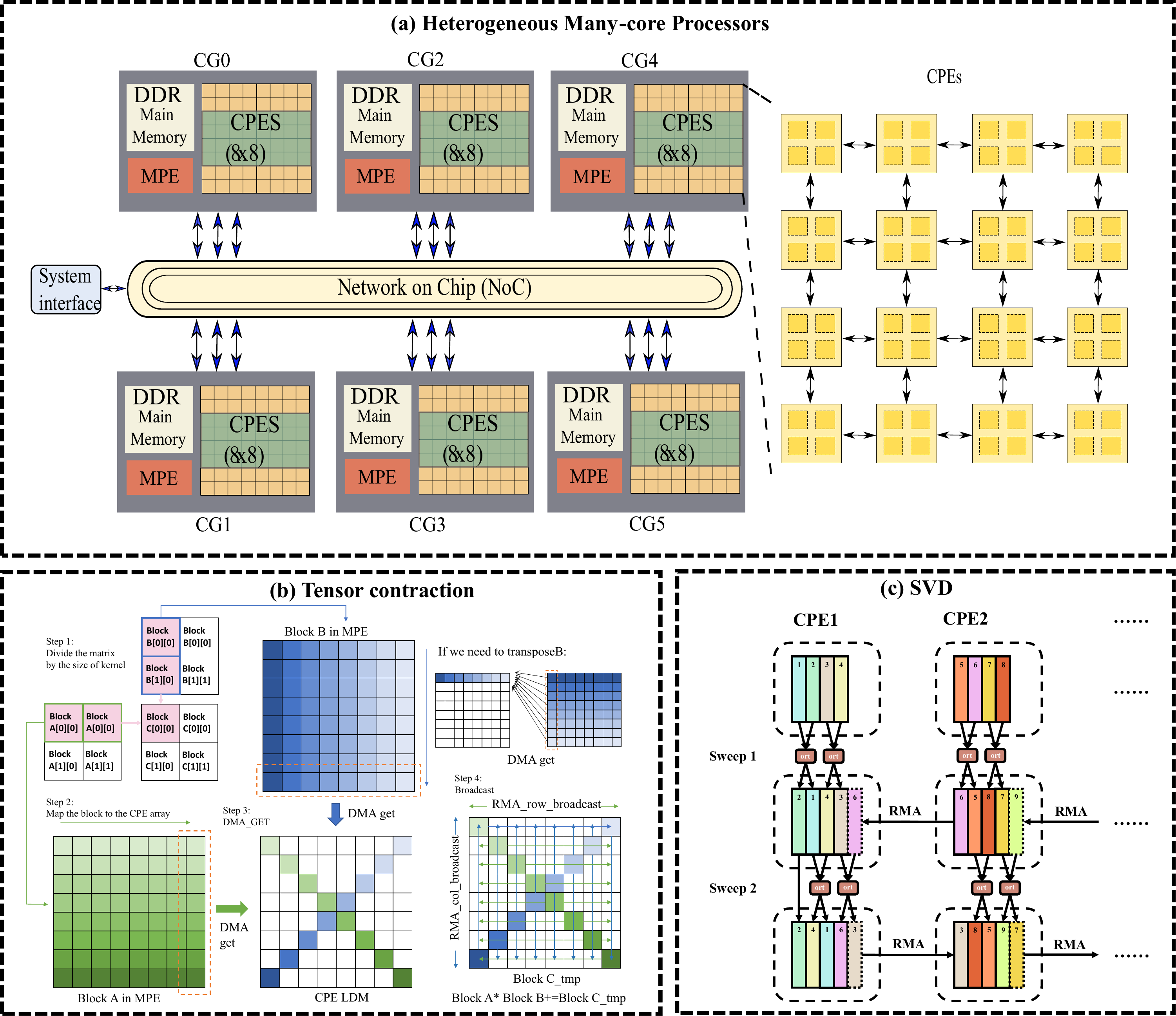}
	\caption{Algorithm details for linear algebra routines. (a) Architecture of the SW26010Pro processor. (b) Matrix multiplication on the Sunway many-core processor. (c) One-sided Jacobi SVD algorithm on the Sunway 
		many-core processor. }
	\label{fig:sw-zgemm-svd}
\end{figure*}

\begin{figure}
	\centerline{\includegraphics[width=0.9\columnwidth]{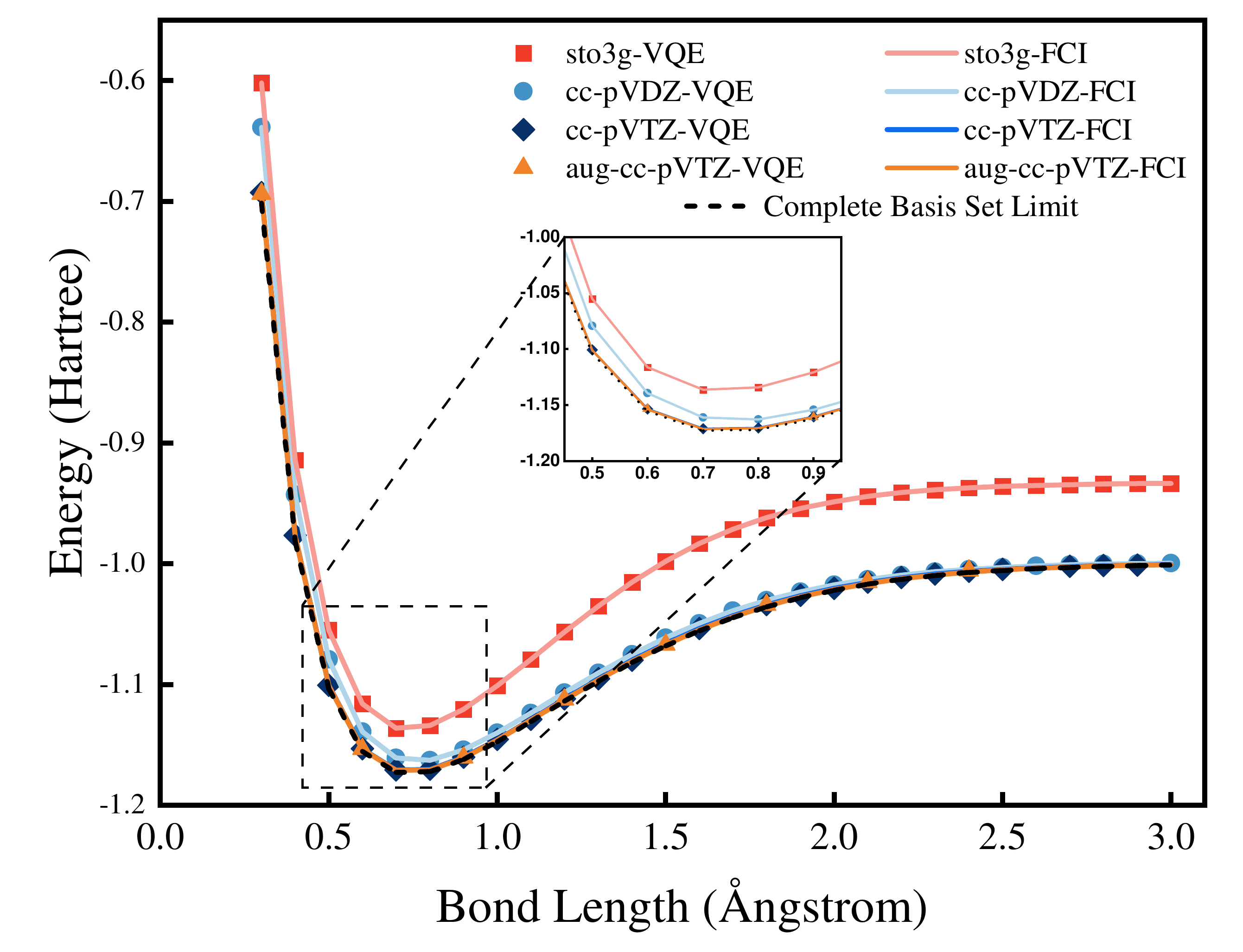}}
	\caption{Potential energy curves \fanyi{in unit Hartree} of the hydrogen molecule computed with UCCSD. The basis sets are STO-3g, cc-pVDZ, cc-pVTZ and aug-cc-pVTZ , \fanyi{which correspond to 4, 20, 56 and 92 qubits, respectively.} \fanyi{The results of full configuration interaction calculations at the complete basis set limit are provided for comparison.}}
	\label{fig:h2}
\end{figure}

\begin{figure}
	\centering
	\includegraphics[width=0.98\columnwidth]{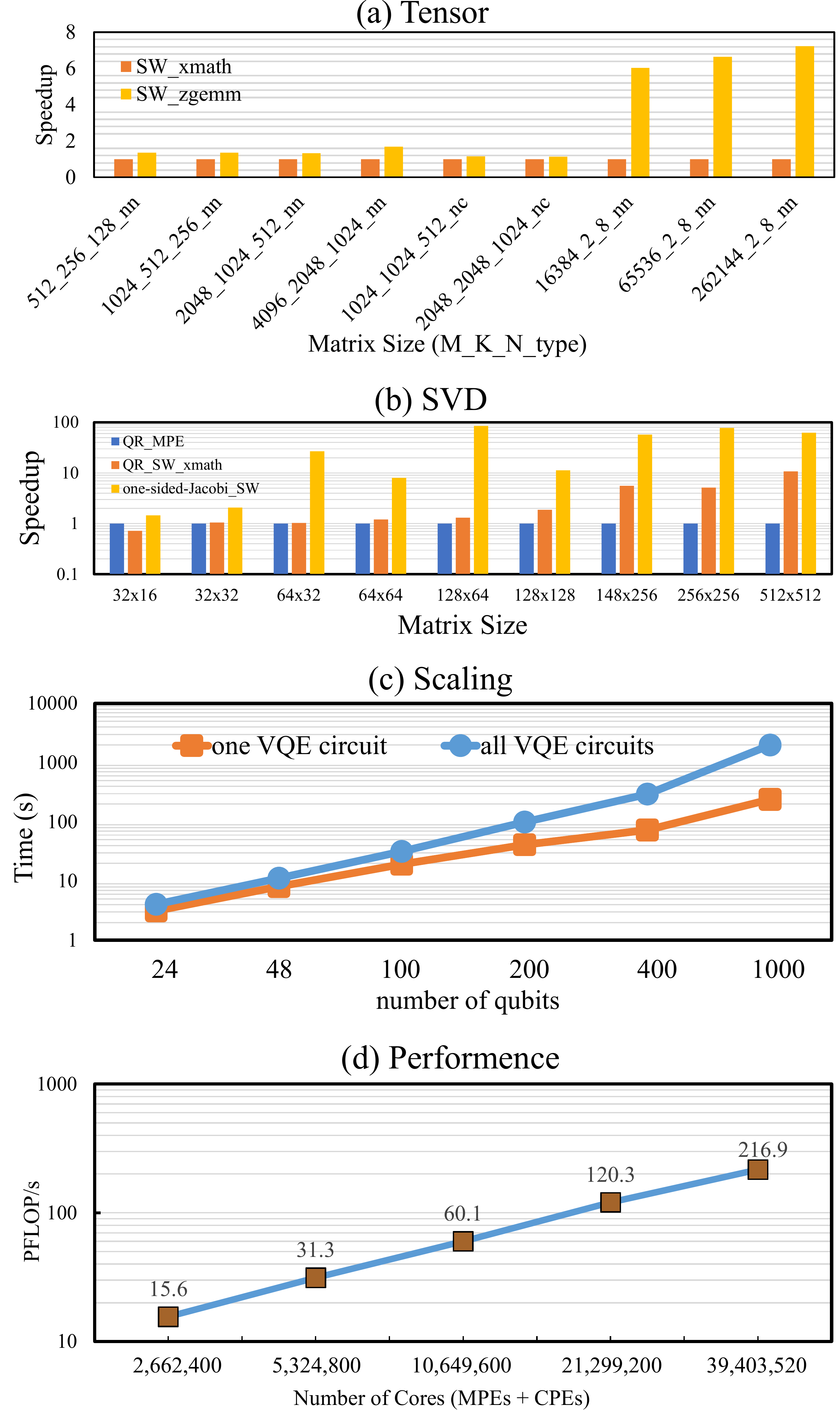}
	\caption{Performance results of linear algebra routines. (a) The performance comparison for tensor contraction with respect to the matrix size. (b) The performance comparison for SVD with respect to the matrix size, \shang{which are evaluated on one CG that contains 1 MPE and 64 CPEs.} (c) The computational time of the hydrogen chain with the MPS-VQE simulator. The blue line refers to the computational time for all circuits with 512 processes. (d) The performance~(PFLOP/s) and strong scaling of the MPS-VQE simulator integrated with DMET on the new-generation Sunway supercomputer.  }
	\label{fig:scaling}
\end{figure}

\begin{figure}
	\centering
	\includegraphics[width=0.98\columnwidth]{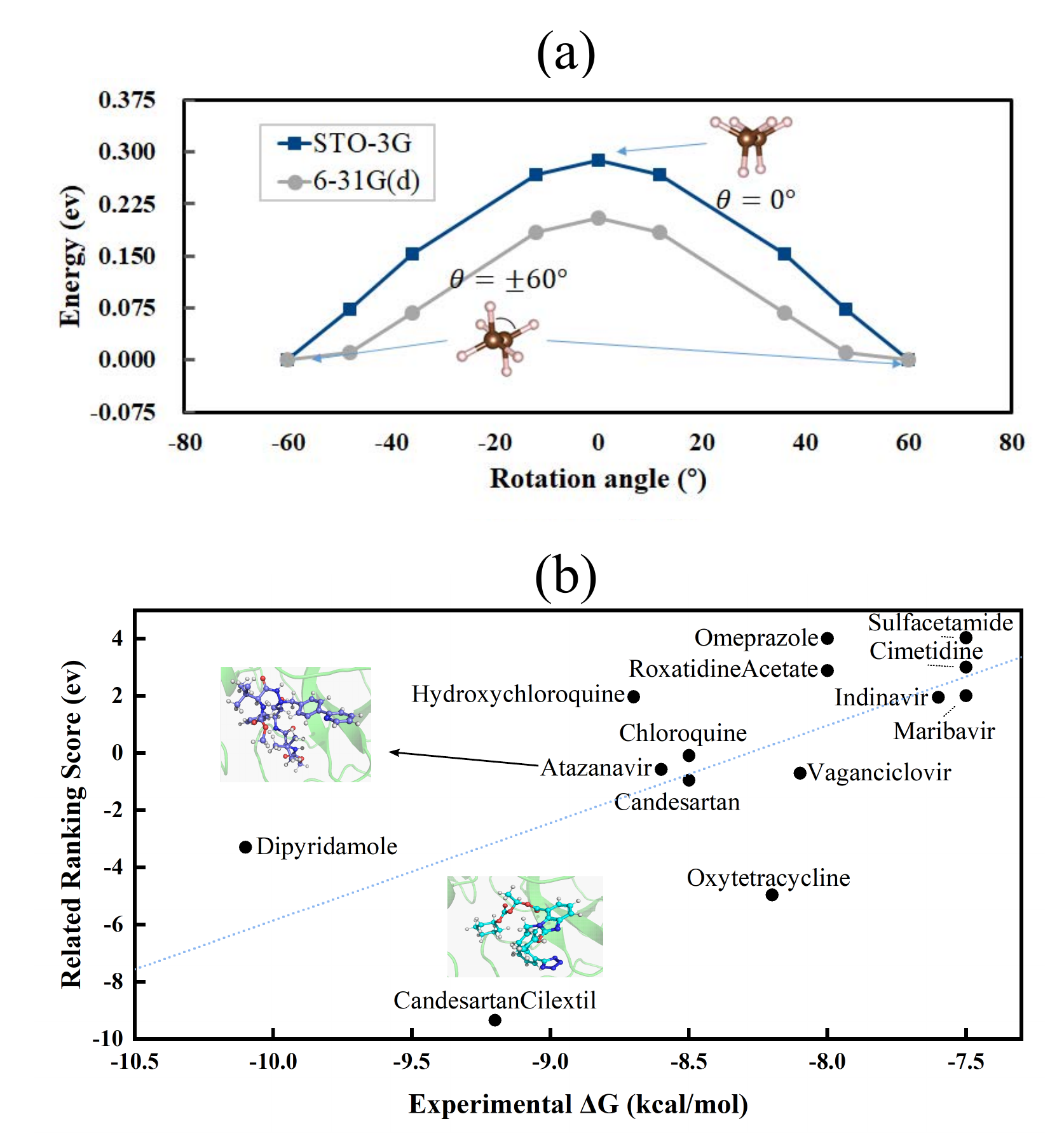}
	\caption{Simulated results for chemical applications. (a) Torsional barrier of the ethane molecule simulated with MPS-VQE \fanyi{using STO-3g~(32 qubits) and 6-31G(d)~(12 qubits using a (6e,6o) active space) basis set}.  (b) Binding energy ranking score versus experimental binding free energies. The overall R$^2$ value for all points is 0.44. The results are computed with MPS-VQE integrated with DMET.}
	\label{fig:ethane}
\end{figure}

\begin{figure}
	\centering
	\includegraphics[width=0.98\columnwidth]{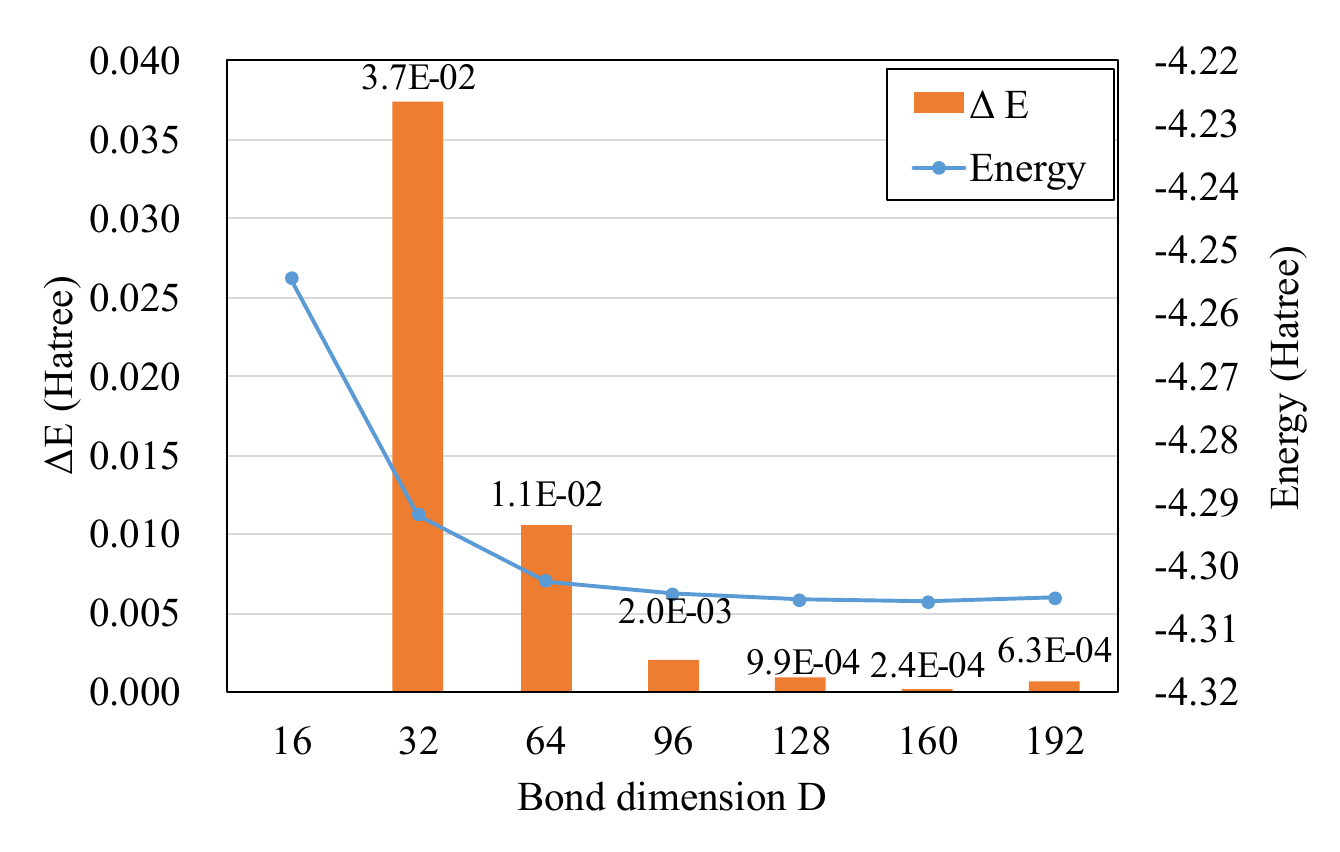}
	\caption{The MPS-VQE optimized energies of H$_8$ molecule using different bond dimension settings. The energy different $\Delta E$ is calculated by $\Delta E=|E_{D_i}-E_{D_{i+1}| }$, where $D_{i} \in \{16, 32, 64, 96, 128, 160, 192\}$.}
	\label{fig:mps-d-benchmark}
\end{figure}

\begin{figure}
	\centering
	\includegraphics[width=0.80\columnwidth]{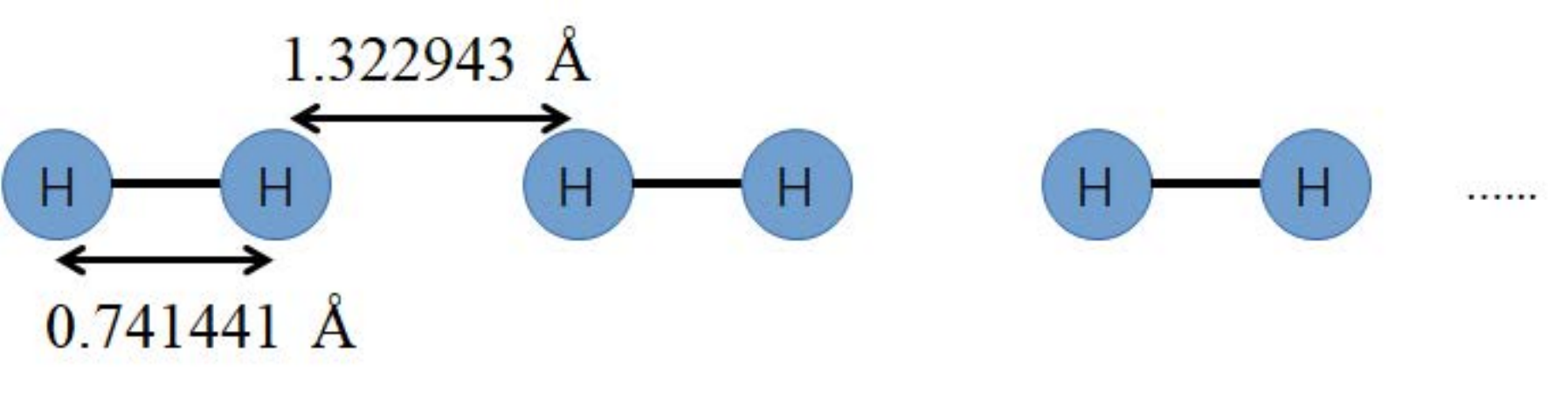}
	\caption{\newshang{The geometry of the one-dimensional hydrogen molecule chain.} The hydrogen atoms are placed with alternate bond lengths of 0.741441 \r{A}ngstrom and 1.322943 \r{A}ngstrom.}
	\label{fig:h2-mol-chain}
\end{figure}